\documentclass[english,aps,graphicx,amsmath,showpacs,prl,preprint]{revtex4}
\usepackage[T1]{fontenc}
\usepackage[latin1]{inputenc}
\usepackage{babel}
\usepackage{graphics}

\makeatletter

\providecommand{\LyX}{L\kern-.1667em\lower.25em\hbox{Y}\kern-.125emX\@}

\makeatother
\begin{document}

\title{Public-channel cryptography using chaos synchronization}

\author{Einat Klein\( ^{1} \), Rachel Mislovaty\( ^{1} \), Ido Kanter\( ^{1} \) and Wolfgang Kinzel\( ^{2} \)}

\affiliation{\( ^{1} \)Minerva Center and Department of Physics,
Bar-Ilan University, Ramat-Gan, 52900 Israel,}

\affiliation{\( ^{2} \)Institute for Theoretical Physics,
University of W\"urzburg, Am Hubland 97074 W\"urzburg, Germany}

\begin{abstract}
We present a key-exchange protocol that comprises two parties with
chaotic dynamics that are mutually coupled and undergo a
synchronization process, at the end of which they can use their
identical dynamical state as an encryption key. The transferred
coupling-signals are based non-linearly on time-delayed states of
the parties, and therefore they conceal the parties' current state
and can be transferred over a \emph{public} channel.
Synchronization time is linear in the number of synchronized
digits $\alpha$, while the probability for an attacker to
synchronize with the parties drops exponentially with $\alpha$. To
achieve security with finite $\alpha$ we use a network.
\end{abstract}


\maketitle

The idea of using chaos for secure communication systems has been
the focus of many research projects in the last few years
\cite{chaoscrypto1,chaoscrypto2,chaoscrypto3,chaoscrypto4,chaoscrypto5}.
Although chaotic systems are linearly unstable and unpredictable,
they can synchronize \cite{Pikovsky}, which makes them promising
candidates for constructing  cryptographic systems. However, in
chaotic cryptographic systems until now, the partners had to agree
on some secret parameter, which somehow had to be transferred
privately. After this private agreement, the two chaotic systems
synchronize by exchanging signals over a public channel, which
then can be used to conceal the message. However, modern
cryptographic protocols construct secret keys over a \emph{public}
channel. Here we investigate whether such a key exchange is
possible using chaotic synchronization.

We present a chaotic system that constructs a \emph{secret} key
using a \emph{public} channel, i.e., a cryptographic key-exchange
protocol base on chaotic synchronization. In our approach, each
party uses a chaotic system. Both of the systems are coupled by
exchanging public signals in order to synchronize. Soon after
synchronization one of the chaotic variables is used as an
encryption key. Although an eavesdropper, listening to the
communication channel, knows all the details of the systems
including the values of the parameters as well as the signals
transmitted, he does not manage to construct the secret key.

The first method in constructing a secret key over a public
channel was developed in 1976 by Diffie and Hellmann. This method
is based on number theory, its complexity is polynomial with the
size of the key, and is the basis of almost all modern encryption
protocols. In view of applications in communication by electronic
circuits or lasers \cite{chaoscrypto5}, it would be useful to find
cryptographic systems based on continuous signals, and with linear
complexity. Our method uses chaotic ordinary differential
equations which are coupled by a few of their internal variables.
For a cryptographic application we have to add two additional
ingredients to the chaotic system: nonlinearity and time delay of
the transmitted signals.

The coupling leads to synchronization of the two ODEs, and the
partners use one of the variable at some predefined time $t$ as
the secret encryption key. Any eavesdropper is at a disadvantage
in that he can only listen but cannot influence the
synchronization process of the two partners. Therefore an attacker
cannot find the secret key by unidirectional coupling
\cite{Michal}. The difference between bidirectional coupling of
the partners and unidirectional coupling of an attacker is the
essence of our cryptosystem.

Because the coupling signals are transferred publicly, they must
be sophisticated enough to hide the state of the systems. We used
coupling signals that are based non-linearly on time-delayed
values of the systems, and therefore conceal the system's current
state, while still enabling synchronization. Time-delayed coupling
has been recently studied \cite{TimeDelay1,TimeDelay2} and is also
observed in systems such as coupled lasers and spiking neurons.

Let us now describe the system in more detail. Consider two Lorenz
systems, $A$ and $B$, coupled by their $x$-value:

\begin{eqnarray}
\frac{dx_{A}}{dt}=10(y_{A}-x_{A})+K[f_{B}(t)-f_{A}(t)]~~~~~~~~~~~~~~~~~~~~~ &  & \label{eq_one}\\
\frac{dx_{B}}{dt}=10(y_{B}-x_{B})+K[f_{A}(t)-f_{B}(t)]~~~~~~~~~~~~~~~~~~~~~&  & \nonumber\\
\nonumber\end{eqnarray}

\vspace{-1.2 cm}
\begin{eqnarray}
\frac{dy_{A}}{dt}=28x_{A}-y_{A}-x_{A}z_{A}~~~\frac{dy_{B}}{dt}=28x_{B}-y_{B}-x_{B}z_{B}~~~~ &  & \nonumber\\
\frac{dz_{A}}{dt}=x_{A}y_{A}-\frac{8}{3}z_{A}~~~~~~~~~~~~\frac{dz_{B}}{dt}=x_{B}y_{B}-\frac{8}{3}z_{B}~~~~~~~~~~~~\nonumber\\
\nonumber\end{eqnarray}

\noindent where $K$ is the coupling strength between the two
systems, $f(t)$ is a non-linear function based on $x$ at previous
time steps: $f(t)=f(x(t-\tau_1),x(t-\tau_2)...)$.

\begin{figure}
\vspace{-0.4 cm} {\centering
\resizebox*{0.5\textwidth}{0.35\textheight}
{{\includegraphics{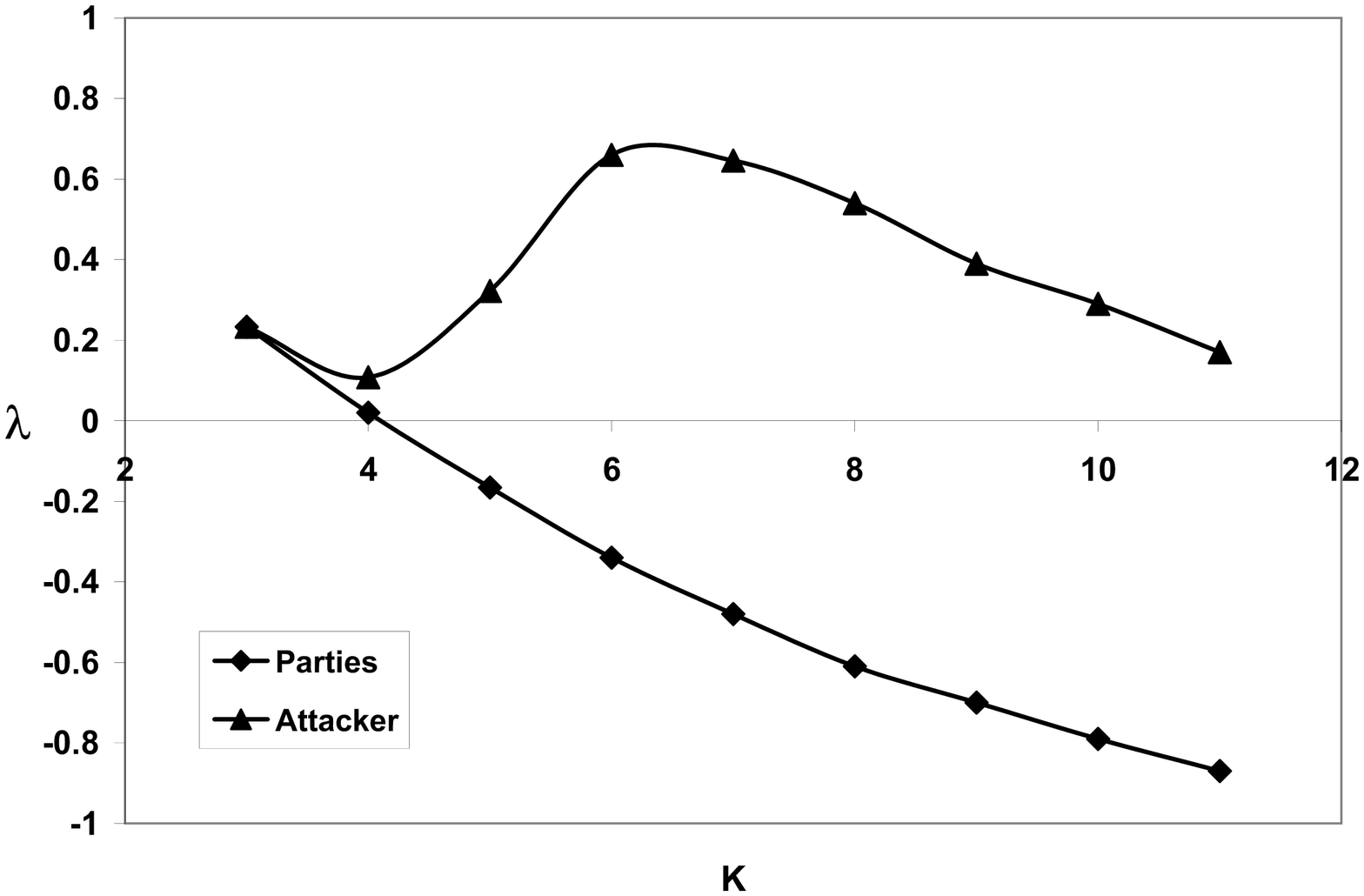}}}
\par}

\vspace{-0.9 cm}\caption{\label{LE} The conditional Lyapunov
exponent versus K, for parties (squares), and the attacker
(triangles). Inset: The probability of synchronization vs. K. For
both graphs $\tau_{1}=0.1$, $\tau_{2}=0.05$ and $A=0.3$. }
\end{figure}

\begin{figure}
\vspace{0.1 cm} {\centering
\resizebox*{0.5\textwidth}{0.35\textheight}
{{\includegraphics{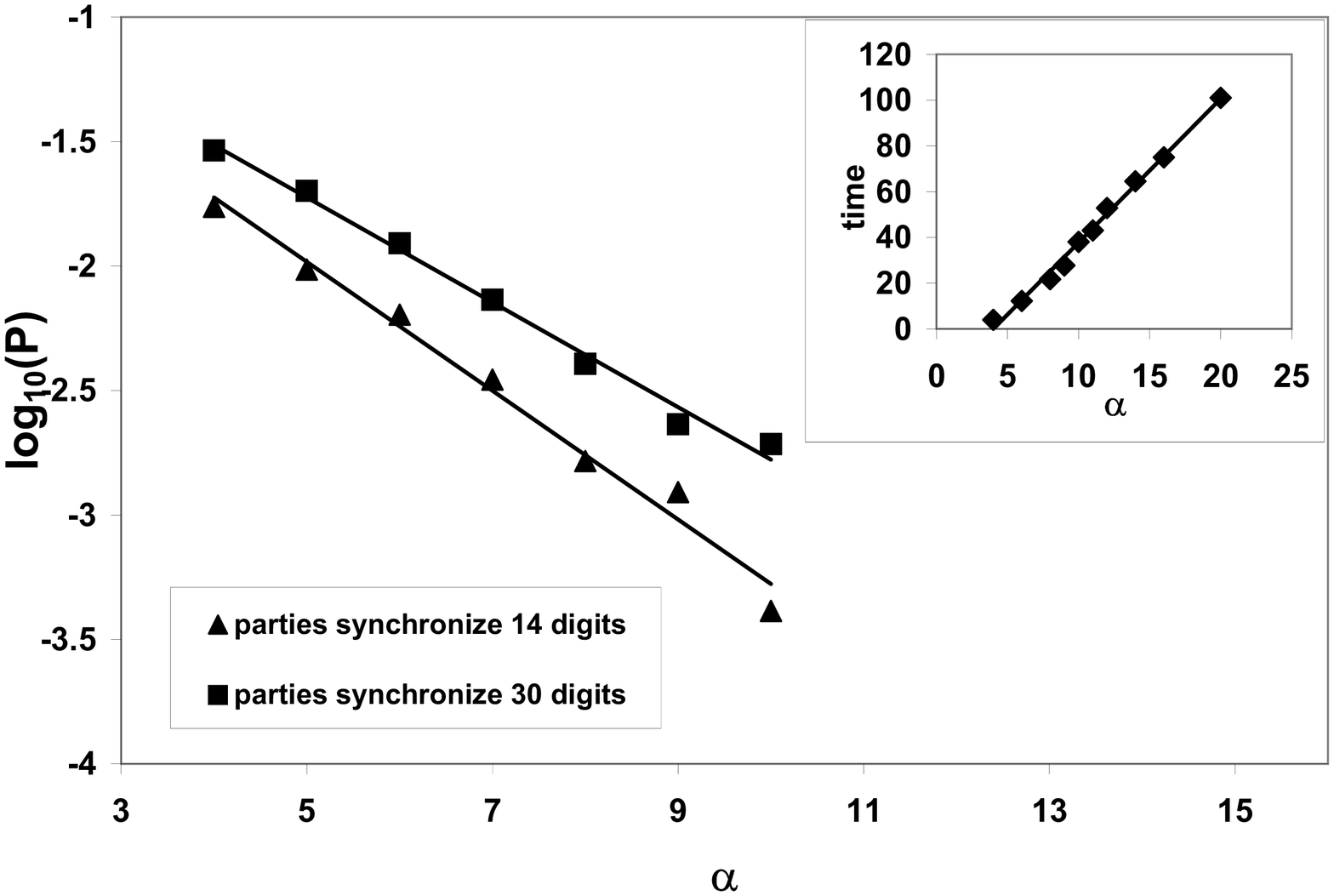}}}
\par}

\vspace{-1.4 cm}\caption{\label{N1} A semi-log plot of the
attacker's success probability versus $\alpha$, for parties
synchronizing 14 digits (triangles), and 30 digits (squares). The
parties use $K=8$ and the attacker uses $K=14$. Inset: The
parties' synchronization time vs. number of synchronized digits,
with $K=8$. For both graphs $\tau_{1}=0.1$, $\tau_{2}=0.05$ and
$A=0.3$. }
\end{figure}

Each party initializes its variables with secret random values.
Use of $x(t- \tau)$ as the coupling signal is not secure,
therefore we suggest using a nonlinear function of the variable
$x$ at previous time steps, $f(t)$ as described above. Which
nonlinear function $f$ should be used? On one hand, $f(t)$ should
enable synchronization. If we choose a signal that is too far from
the $x$ value, e.g. $x(t-\tau)$ for a large time delay $\tau$, the
systems will not synchronize. On the other hand, if we choose a
function $f$ which is linear in $x(t-\tau)$ , it will be easy to
reveal the state of the system. Therefore we add a small
perturbation to the main signal, constructed from two delayed
values $x(t_{1})$ and $x(t_{2})$, where $t_{1}=t-\tau_{1}$ and
$t_{2}=t-\tau_{2}$ :

\vspace{-0.4 cm}
\begin{equation}
f(t)=x(t_{1})+sgn(x(t_{1}))A(x(t_{1})-x(t_{2}))^2\label{eq_five}
\end{equation}

\noindent where $sgn(x(t_1))$ ensures an average mean for the
perturbation around $x(t_1)$.

Our numerical simulations show that synchronization is possible up
to the critical value $A_{c}\sim 0.36$, for $\tau_{1}=0.1$ and
$\tau_{2}=0.05$. When approaching this value there is a
probability close to 1 that the systems' variables diverge.

We find that with a time delayed-signal, there is a nonzero
probability for synchronization only in a limited range $K_{min}<
K < K_{max}$. For the parameters of Fig. \ref{LE}, for instance,
$K_{min}\sim 4$ and $K_{max}\sim 11.5$. Small values of the
coupling strength $K$ are  too weak to achieve synchronization. On
the other hand, large values of $K$ lead to a nonzero probability
that the variables of the Lorenz systems $(x,y,z)$ diverge. Above
$K_{max}$ all initial states diverge. Similarly, the time delay
must not be too large to achieve synchronization. We find a
maximal value of
 $\tau_i < \tau_{max} \sim 0.12$. Synchronization is therefore possible
only in a limited range of model parameters. It turns out that
this is essential for the cryptographic application.


In the following we consider an attacker $E$ who knows all the
details of the model and listens to any communication between the
parties $A$ and $B$. The first attack strategy we discuss is an
attacker who uses the same Lorenz system as the two parties,
follows their steps throughout the process, and also uses the same
signal $f_A(t)$ in order to synchronize.

\begin{equation}
\frac{dx_{E}}{dt}=10(y_{E}-x_{E})+K[f_{A}(t)-f_{E}(t)]\label{eq_attacker}\\
\end{equation}

We name this attack the "regular following attack" (RFA). The RFA
may use a larger coupling strength $K$ to increase his tracing
steps \cite{Pikovsky}.

Therefore, we have to investigate the behavior of the
bidirectionally coupled $A$/$B$ system and the unidirectionally
coupled $A$/$E$ system as the function of the coupling strength
$K$. A quantitative measure of synchronization is given by the
conditional Lyapunov exponents (CLE). Synchronization is possible
only if all the CLE of the systems are negative
\cite{chaoscrypto1}. Fig. \ref{LE} shows the largest CLE of the
parties (squares), and the RFA attacker (triangles).

The standard technique for measuring the CLE is not applicable in
our case, since one has to approximate the time delayed values,
for the parties and the attacker. To overcome this difficulty we
use the following "self-consistent" procedure, which is a
variation on the standard method. The parties start from a point
on the attractor, with a small distance $d_{0}=10^{-8}$ between
them. We assume a CLE and then generate the appropriate time
delayed values. Given the time delayed values and the current
state of the parties, the CLE can now be calculated. The correct
CLE is the one for which the measured CLE matched the assumed CLE
used for generating the time delayed values. A similar procedure
is used for estimating the CLE of the attacker.

Fig. \ref{LE} indicates that the CLE of the parties is negative
for $5 \le K \le 11$, while the CLE of the attacker is positive in
this regime. Note that for $K >11.5$, the $x$ values (and also $y$
and $z$) of the parties and the attacker diverge. In practice we
noticed that the the most successful attacker is the one using
$K=14$, while the attacker bounds his x-values to $|x|<22$ so as
not to diverge.

Fig. \ref{N1} displays a semi-log plot of the RFA attacker's
success probability vs.$\alpha$, the number of digits he manages
to synchronize, when the parties are synchronized by 14 digits
(triangles) and 30 digits (squares). His success probability drops
exponentially with $\alpha$, whereas the synchronization time of
the parties grows linearly with $\alpha$, as shown in the inset of
Fig. \ref{N1}. Therefore the parties can still use most of their
digits for the encryption key.

These results show that chaotic ODEs can be used to generate a
secret key over a public channel. There is an interplay between
either the positive CLE or the lack of an attractor (the
divergence of the parameters of the attacker) and the security of
our cryptosystem.

We have seen that the attacker cannot synchronize with the two
parties. However, he may try to analyse the exchanged signals
$f_A(t)$ in order to calculate the variable $x_A(t)$. The
following attack, named the embedded signal attack (ESA), tries to
analyse the transmitted signal by embedding the signal in a space,
defined by signals transmitted in different time steps
\cite{Pikovsky}.

A space $F=\{f(t),f(t'),f(t")...\}$ is defined, which uses a
sequence of $f$ values from different time steps $t$, $t'$, etc.
The attacker tries to map the F-space to corresponding $x$-values
of the system. If for a small window in F-space there is a small
corresponding range of $x$-values, then this  mapping is possible.
Yet if the distribution of $x$-values corresponding to a small
window in F-space is wide, then $x$ is not uniquely defined and
mapping is not possible.

\begin{figure}
{\centering \resizebox*{0.5\textwidth}{0.3\textheight}
{{\includegraphics{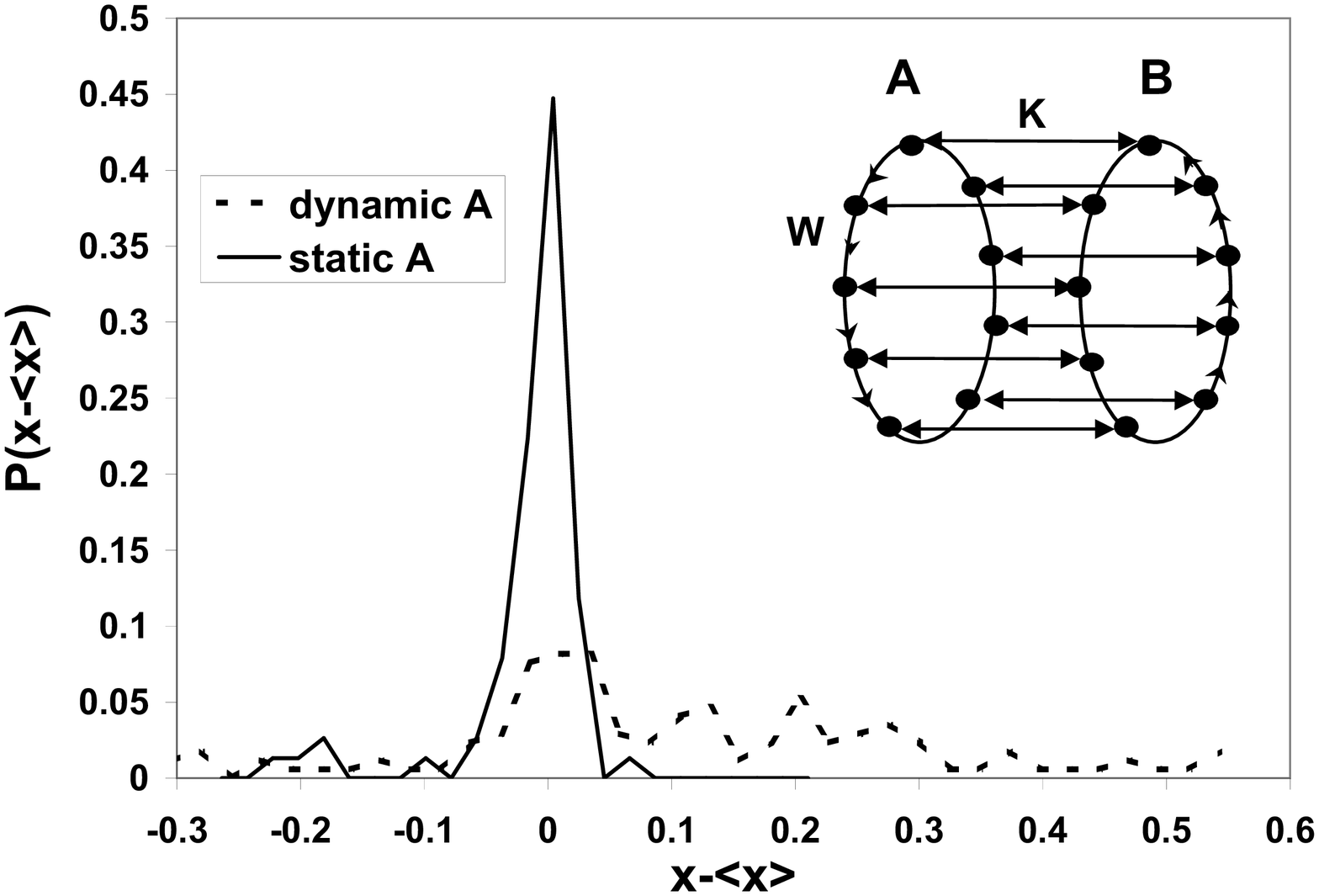}}}
\par}

\vspace{-0.9 cm}\caption{\label{Adynamic}The ESA attack on one
node, N=1. The graph shows the distribution of $x$ values for a
window in F-space, $F\{f(t),f(t-0.3),f(t-0.9)\}$, with window edge
of 0.02. For dynamic $A$ as defined in Eq. \ref{eq_A} (dashed
line) and static $A=0.01$ (black line). Inset: Schematic figure of
the two coupled cyclic networks. Each node represents a Lorenz
system and is coupled to the preceding node, and to a parallel
node in the other network.}
\end{figure}

Fig. \ref{Adynamic} shows the distribution of $x$ values for a
window of size $0.02$ in F-space, $F\{f(t),f(t-0.3),f(t-0.9)\}$.
The distribution of $x$-values is peaked, therefore the ESA is
successful for a \emph{finite} $\alpha$. However, for
\emph{infinite} $\alpha$ one has to decrease the window's size
accordingly. The probability of finding a point in the dynamics
belonging to such a tiny window decreases exponentially with
$\alpha$, and therefore this attack reduces to a brute force
attack. Hence, the presented cryptosystem is also robust against
the ESA attack.

In order to increase the key space and to decrease the precision
of the calculation we investigated an extension of the system to a
network of $N$ Lorenz equations. Now each party has a ring of
Lorenz systems which are coupled as shown in Fig.  \ref{Adynamic}.
We tried other topologies as well, but it turned out that the
cyclic network yields the highest security. The network generates
a key of size $\alpha N$ and the security is a function of network
size $N$.

The two cyclic networks $A$ and $B$ are coupled: each node is
coupled to a parallel node in the other network and to its
preceding neighbor by its $x$ value. The two networks exchange N
signals, ${f(x^{i})}$ $i=1...N$, at every time step, and use the
following dynamics:

\begin{eqnarray}
\frac{dx_{A}^{i}}{dt}=10(y_{A}-x_{A})+K[f^{i}_{B}(t)-f^{i}_{A}(t)]~~~~~~~~~ &  & \label{eq_N}\\
+W[g^{i+1}_{A}(t)-g^{i}_{A}(t)]~~~~~~&  & \nonumber\\
\nonumber\end{eqnarray}

\noindent and similarly for system $B$, where $f^{i}(t)$ is given
by Eq. \ref{eq_five} for node $i$ in the network. $K$ is the
strength of the coupling between systems $A$ and $B$ and $W$ is
the strength of the inner coupling ("weights"). For simplicity we
use $g(x)=f(x)$ (see Eq. \ref{eq_five}) with $A=0.1$.

Our simulations show that the two systems reach a state of
synchronization in which, although the values of the two networks
are identical, there are no clusters among the nodes of each
network and they are de-synchronized (if the inner coupling
strength W is not too strong). Note that although the systems now
exchange N signals, the signal's size is small because we use a
small $\alpha$.

If the parties synchronize a finite number of digits, another
modification must be made in the model to increase security. The
attacker's probability of divergence grows with $A$, therefore we
can enhance the security even further by using a $dynamic$
amplitude $A$ in Eq. \ref{eq_five}, in the following way:

\begin{equation}
A(t)=\frac{1}{B|f_{A}(t)-f_{B}(t)|^{\rho}+C}\label{eq_A}
\end{equation}

\noindent where $B$ and $C$ can be constants, or stochastic
numbers following a known protocol. At first $A$ is relatively
low, so that the parties will start coming closer. Gradually they
get closer and $A$ grows so that synchronization becomes more
difficult. Because the attacker's probability to diverge is
higher, using a dynamic pre-factor $A$ affects him much more than
it does the parties; even if he bounds his values $(x,y,z)$ so as
not to diverge, his success probability is greatly reduced.

\begin{figure}
{\centering \resizebox*{0.5\textwidth}{0.3\textheight}
{{\includegraphics{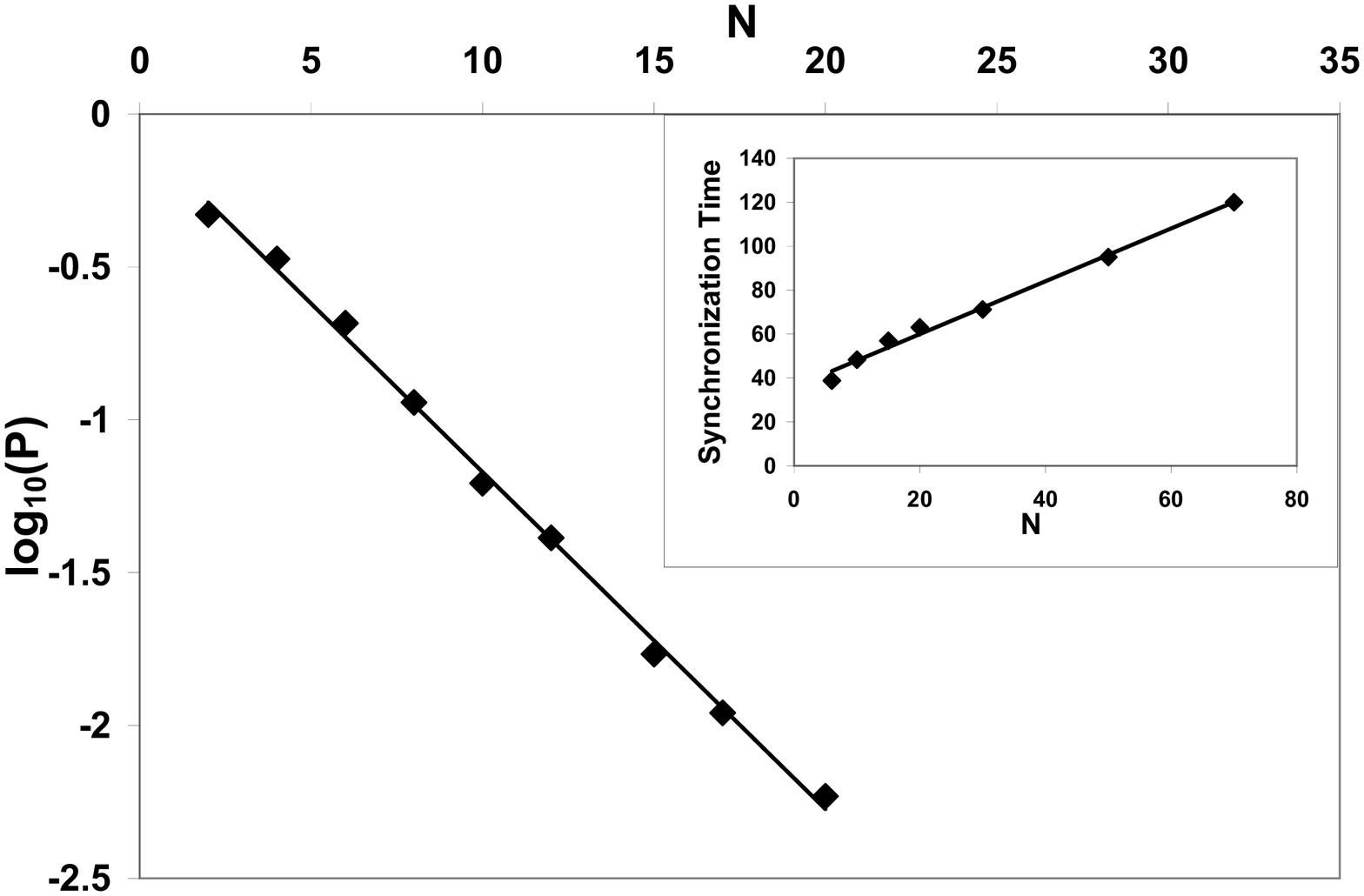}}}
\par}

\vspace{-0.9 cm}\caption{\label{lnP}Semi-log plot of the
probability of the RFA attacker to synchronize one node, versus N.
Inset: The parties' synchronization time versus N. For both graphs
the parties use K=8 and W=2  and the attacker uses K=14 and W=2,
$\rho =1.5$, $B=200$ and $C$ is randomly chosen in the range
$[3,4]$.}
\end{figure}

The RFA attacker tries to synchronize with the parties. When is he
considered successful? When he synchronizes all the nodes
completely? Synchronizing only part of them is probably enough. We
set a very soft criterion and considered a successful attacker one
who manages to synchronize at least one node by only 4 digits,
while the parties synchronize $all$ the nodes by 7 digits. Albeit
the soft criterion, we observed that the probability for an
attacker to succeed decays exponentially with N, as demonstrated
in Fig. \ref{lnP}. The parties' synchronization time on the other
hand, scales with $N$, as displayed in the inset of Fig.
\ref{lnP}.

Using a network increases the security against RFA. Every Lorenz
system in the attacker's network has a probability to diverge.
When using a network, if there is a node that diverges, it also
affects its neighbors and they too start to diverge. It is like
damage-spreading. The attacker finds it difficult to prevent this
from occurrance because if he cuts out even one diverging node, he
is left with an open chain.

The ESA attacker is relevant only to the case of finite $\alpha$.
We find that using a dynamic $A$ as defined in Eq. \ref{eq_A}
increases the security against ESA even for $N=1$. Fig.
\ref{Adynamic} shows the distribution of $x$ values for a window
in F-space, $F\{f(t),f(t-0.3),f(t-0.9)\}$. When using a static
$A$, the distribution of $x$-values is peaked, therefore the ESA
is successful for a finite $\alpha$. However when using a dynamic
stochastic $A$, the distribution of $x$-values corresponding to a
small window in F-space is wide. Because $A$ is dynamic and
stochastic, there exist many close trajectories of $f$ that lead
to different $x$-values.

Note that another type of attack suggested for cryptographic
systems based on synchronization of neural networks \cite{KKK} is
irrelevant to this system. The 'Majority Attack' is based on an
ensemble of cooperating attackers \cite{Majority}. Cooperating
attackers are ineffective here because of the linear instability
of the dynamics.

To conclude, the ability of two chaotic systems to synchronize
when coupled by a time-delayed signal is used to create a
cryptographic system. The signals do not reveal the state of the
system, yet still enable synchronization. One coupled Lorenz pair
is secure when controlling $\alpha$. A secure cryptographic system
is constructed by weakly coupling N Lorenz systems, enabling the
use of less precision in the calculations. Several factors
contribute to the security of this system: the linear instability
of the dynamics, the fact that the two parties are mutually
coupled while the attacker is one-way coupled, and the structure
of the network which allows individual defects to affect the
entire system.

Fruitful discussions with Arkady Pikovsky are acknowledged. The
research of I.K is partially supported by the Israel Academy of
Science.

\end{document}